\documentclass[epj]{svjour}
\newcommand{\rmd}{\mathrm{d}}
\newcommand{\rmi}{\mathrm{i}}
\newcommand{\rme}{\mathrm{e}}
\usepackage{graphics}
\begin{document}
\title{Coherent XUV generation driven by sharp metal tips photoemission}

\author{M. F. Ciappina\inst{1}\thanks{\emph{Present address:} Max-Planck-Institut f\"ur Quantenoptik, Hans-Kopfermann-Strasse 1, D-85748 Garching, Germany} \and J. A. P{\'e}rez-Hern{\'a}ndez\inst{2}
\and T. Shaaran\inst{3} \and M. Lewenstein\inst{4,5} 
%\and M. Kr{\"u}ger\inst{6,7} \and P. Hommelhoff\inst{6,7}
% etc
}       
\institute{Department of Physics, Auburn University, Auburn, Alabama 36849, USA \and Centro de L\'aseres Pulsados (CLPU), Parque Cient\'{\i}fico, 37185 Villamayor, Salamanca, Spain \and CEA-Saclay, IRAMIS, Service des Photons, Atomes et Molecules, F-91191 Gift-sur-Yvette, France \and ICFO-Institut de Ci\'encies Fot\'oniques, Mediterranean Technology Park, 08860 Castelldefels (Barcelona), Spain \and ICREA-Instituci\'o Catalana de Recerca i Estudis Avan\c{c}ats, Lluis Companys 23, 08010 Barcelona, Spain 
%\and Department of Physics, Friedrich-Alexander-Universit\"at Erlangen-N\"urnberg, Staudtstr.~1, D-91058 Erlangen, Germany, \and Ultrafast Quantum Optics Group, Max-Planck-Institut f\"ur Quantenoptik, Hans-Kopfermann-Str.~1, D-85748 Garching bei M\"unchen, Germany
}
\date{Received: date / Revised version: date}
\abstract{
It was already experimentally demonstrated that high-energy electrons can be generated using metal nanotips as active media. In addition, it has been theoretically proven that the high-energy tail of the photoemitted electrons is intrinsically linked to the recollision phenomenon. Through this recollision process it is also possible to convert the energy gained by the laser-emitted electron in the continuum in a coherent XUV photon. It means the emission of harmonic radiation appears to be feasible, although it has not been experimentally demonstrated hitherto till now. In this paper, we employ a quantum mechanical approach to model the electron dipole moment including both the laser experimental conditions and the bulk matter properties and predict is possible to generate coherent UV and XUV radiation using metal nanotips as sources. Our quantum mechanical results are fully supported by their classical counterparts.
 % end of PACS codes
} %end of abstract
\maketitle
\section{Introduction}
\label{intro}

One of the most noticeable examples of the nonlinear process in the laser-matter interaction is the generation of coherent radiation from the ultraviolet (UV) to extreme ultraviolet (XUV) range. In particular, when an intense laser beam is focused into a gas jet of atoms or molecules the conversion of a photon of determined energy in a one with much higher energy could occur~\cite{McPherson1987,Huillier1991}. This phenomenon is known as high-order harmonic generation (HHG). The interest of HHG resides in the fact that it represents one of the most reliable paths to generate coherent UV to XUV light. Furthermore, HHG with basis on atoms and molecules has proven to be a strong source for the generation of attosecond (as) pulses trains~\cite{antoine}, that can be temporally confined to a single XUV as pulse, nowadays with repetition rates in the range of the kHz~\cite{Scrinzi2004,hhg1khz,hhg3khz,amelle}. HHG has a set of remarkable properties which can as well be employed, for instance, to extract temporal and spatial information with both attosecond and sub-{\AA}ngstr\"om resolution on the generating system~\cite{manfred_rev}, or to dissect the atomic world within its natural temporal and spatial scales~\cite{pacer,rabitt,olga1,olga2,mairesse,power}.

Recently it has been put forward the utilization of {\it alternative} active media to study strong field phenomena, namely metal nanotips~\cite{peter2011}, nanoparticles~\cite{klingnature} or ablation plumes~\cite{ganeev1,ganeev2} (for a recent review see e.g.,~\cite{peter2012}). For instance, due to the interaction of an intense laser pulse with a metallic nanostructure, one can generate very high-energy
electrons. This phenomenon, which is called above-threshold
photoemission (ATP), can be considered as the counterpart of the
above-threshold ionization (ATI), based on atoms and molecules. However,
the underlying physics in ATP is much richer and quite different in its
nature to the ATI (see e.g.~\cite{peter2010,herink}). Several theoretical models have been recently developed and applied in order to understand the experimental outcomes and to guide future measurements~\cite{yalunin,michaelNJP,watcherPRB,ropersAdP}. In addition, it has been demonstrated that solid-state samples can also be employed as generators of high-order harmonic radiation, although the understanding of the HHG phenomenon using bulk matter is at its very beginning~\cite{ghimireexp,ghimiretheory}.

Another analogous process, combining noble gases and bulk matter, is the generation of harmonic radiation using (plasmonically) enhanced fields. The first demonstration of such an effect can be traced back to the experiment of Kim et al~\cite{kim}. In this work was shown it is possible to locally amplify an incoming laser field by employing surface plasmonic resonances. Amplifications in intensity greater than 20 dB can be achieved, manipulating the geometry of the nanostructures~\cite{muhl,schuck}. Consequently, when a low-intensity femtosecond laser pulse couples to the plasmonic mode of a metal nanostructure, it starts a collective oscillation among free electrons within the metal. A region of highly amplified electric field, exceeding the threshold for HHG in gases, can so be achieved. By injection of noble gases surrounding the nanostructure high-order harmonics can be generated. In particular, using gold bow-tie shaped nanostructures Kim et al~\cite{kim} have demonstrated that the initially modest laser field can be amplified sufficiently to generate high energy photons, reaching the XUV regime, and, in addition, the radiation generated from the enhanced laser field, localized at each nanostructure, acts as a point-like source, enabling collimation of this coherent radiation by means of constructive interference (see~\cite{kim} for more details). Recently there has been considerable theoretical work looking at HHG driven by spatially nonhomogeneous fields~\cite{husakou,yavuz,ciappi2012,tahir2012,ciappi_opt,ciappiAdP,miloAdP,joseprl2013,yavuz2013,luo2013,PoP}. However, the initial sudden thrilling about generation of XUV coherent radiation by using plasmonic fields was put in dispute by recent findings~\cite{ropersnat,Kimreply,sivis2013}. Fortunately, other ways to enhance coherent light were examined recently (see e.g. for the production of high-energy photoelectrons using enhanced near-fields from dielectric nanoparticles~\cite{klingnature}, metal nanoparticles~\cite{klingspie,klingprb,lastkling} and metal nanotips~\cite{peter2011,peter2012,peter2010,herink,peter2006,peter2006a,Ropers2007,Barwick2007,Yanagisawa2009,Bormann2010,Park2012}.

In this article we extend previous predictions~\cite{ciappiNanotip} by employing longer laser pulses and adding in more details both about the quantum and classical models. In the next section, Section 2, we describe the quantum mechanical approach, based on the solution of the time-dependent Schr\"o-dinger equation in reduced dimensions and the classical model, based on the Newton equations for one electron moving in an oscillatory electric field. Next, in Section 3, we present numerical results, both using the quantum and classical models, using typical parameters, with emphasis in the sensitivity of the observables to the static electric field. Finally, in Section 4, we close our contribution with our conclusions and a brief outlook. Atomic units are used throughout the article unless otherwise stated.

%We employ available laser source parameters and treat the metal tip with a fully quantum mechanical model within the single active electron approximation (SAE). It is well known the SAE is able to infer the general features behind the laser matter processes, in particular this approximation works correctly to describe the HHG in multielectronic atomic targets and also reproduces satisfactorily experimental results when targets different than atoms are employed (see, e.g.,~\cite{plumes} for HHG in plasma plumes). We do not take into account any collective effect, such as propagation and phase matching, considering it was argued these could play a minor role in the HHG using nanosources (see, e.g.,~\cite{kim}). As it is well known, the main physical mechanism behind the generation of high-order harmonics is the electron recollision and consequently the model used should include it. It was already shown that the recollision mechanism is also needed to describe above-threshold photoemission (ATP) measurements and, considering these two laser-matter phenomena, i.e. the photoemitted electrons and the high-frequency radiation, are physically linked, we could argue that metal nanotips can be used as sources of XUV radiation as well.

\section{Theory}
\label{sec:1}

\subsection{Quantum mechanical approach}
\label{subsec:1}

The theoretical model we use here was already employed for the calculation of the electron photoemission~\cite{peter2006a,peter2012} and high-order harmonic generation (HHG) using ultrashort laser pulses~\cite{ciappiNanotip}, in both cases using metal nanotips as active media.
Consequently, we only give a short description and we emphasize the numerical tools needed to compute the HHG spectra.
In summary, the one dimensional time dependent Schr\"odinger equation (1D-TDSE) is solved numerically within the single active electron (SAE) approximation. The 1D-TDSE can be written as
\begin{eqnarray}
\nonumber
\mathrm{i}\frac{\partial \Psi (z,t)}{\partial t} &=&\mathcal{H}(t)\Psi (z,t)
\label{tdse} \\
&=&\left[ -\frac{1}{2}\frac{\partial ^{2}}{\partial z^{2}}+V_{m}(z)+V_{l}(z,t)\right] \Psi (z,t), 
\end{eqnarray}
where $V_{m}(z)$ defines the potential barrier as 
\begin{equation}
V_{m}(z)=-(W+E_F)
\end{equation}
inside the metal ($z\le 0$) and 
\begin{equation}
V_{m}(z)=-1/(z+\alpha)
\end{equation} 
is the image-charge potential outside ($z> 0$) with $\alpha$ chosen in such a way as to make $V_{m}(z)$ continuous at $z=0$. In addition, 
\begin{equation}
V_l(z,t)=(E_{dc}+E_{L}(t))z 
\end{equation}
is the potential due to the laser oscillating electric field $E_{L}(t)$ including the applied static field $E_{dc}$ that arises due to the DC tip bias voltage.

By using a narrow, few atomic units (au) wide, potential well with variable depth $W+E_F$, where $W$ is the work function and $E_F$ the Fermi energy, we  model the metal surface. This depth and width of the well are chosen in such a way as to match the actual metal sharp tip parameters. In our case we
utilize the parameters for clean gold, i.e. $W=5.5$ eV and $E_F=4.5$ eV, but other typical metals, such as tungsten or others, can be used as well by tunning adequately the values of $W$ and $E_F$. The ground
state of the active electron represents the initial state in the metal nanotip and it is computed by diagonalizing of the discretized hamiltonian $\mathcal{H}(t)$ in absence of the laser potential $V_l(z,t)$.  The infinitely high potential wall on one side and, on the other side a potential step representing the metal-vacuum surface barrier, {\it confine} the electronic wavefunction. Furthermore, the image-force potential gives a smoother shape to the surface barrier potential.
The evanescent part of the electronic wavefunction penetrates into the classically forbidden (vacuum) region. The evanescent behavior for the electronic wavefunction is closely linked to the rescattering phenomenon, which is considered the main responsible of the high-energy tail of the photoelectron spectra and
the high-order harmonic generation.

The oscillating laser electric field is of the form 
\begin{equation}
E_{L}(t)=E_0\,f(t)\sin(\omega t+\phi_{CEP}),
\end{equation} 
where $E_0$, $\omega$ and $\phi_{CEP}$ are the laser electric field peak amplitude, the laser frequency and the carrier-envelope phase (CEP), respectively. The pulse envelope $f(t)$ is chosen to be
\begin{equation}
f(t)=\sin^{2}\left(\frac{\omega t}{2 n_{cy}}\right)
\end{equation}
where $n_{cy}$ is the number of cycles.
The electronic wavefunction $\Psi (z,t)$ of Eq.~(\ref{tdse}) is time propagated using the Crank-Nicolson scheme and the harmonic spectra $D(\omega)$ are retrieved by Fourier-transforming the dipole acceleration $a(t)$ by using:
\begin{equation}
\label{hhg}
D(\omega)=\left| \frac{1}{T_p}\frac{1}{\omega^2}\int_{-\infty}^{\infty}\rmd t\rme^{-\rmi \omega t}a(t)\right|^2,
\end{equation}
where $T_p$ is the total duration of the laser pulse. $a(t)$ of Eq.~(\ref{hhg}) is obtained from the commutator relation,
\begin{equation}
\label{accel1D}
a(t)=\frac{\rmd^{2}\langle z \rangle}{\rmd t^2}=-\langle \Psi(t) | \left[ \mathcal{H}(t),\left[ \mathcal{H}(t),z\right]\right] | \Psi(t) \rangle,
\end{equation}
where $\mathcal{H}(t)$ is the Hamiltonian defined in the Eq.~(\ref{tdse}) (see e.g.~\cite{keitel,schafer} for more details).

\subsection{Classical model}
\label{subsec:2}

In addition to the quantum mechanical approaches the strong field phenomena have been characterized using classical considerations. By employing Newton's second law for an electron moving in a linearly polarized electric oscillating field it is possible to compute the total energy of that electron as a function of the ionization or recombination times. In this approach any influence of the interaction between the laser-ionized electron and the remaining ionic core is completely neglected. This prescription defines what we known as Strong Field Approximation (SFA)~\cite{sfa}. If we include in the classical calculations the static electric field $E_{dc}$ it would be possible to study its influence in the harmonic cutoff values and to characterize the deviations from the conventional simple man's model~\cite{sfa,corkum}. In the latter, valid for atoms and molecules, the harmonic order at the cutoff fulfills $n_{c}=(3.17U_{p}+I_{p})/\omega$, where $n_{c}$ is the harmonic order at the cutoff, $U_{p}$ the ponderomotive energy ($U_p=E_0^{2}/4\omega^2$)~\cite{keitel} and $I_p$ the ionization potential of the atomic or molecular species under consideration. Inserting the values of the peak electric field $E_0$ and laser wavelength and using an equivalent $I_p$ equal to the metal work function $W$ we can corroborate the cutoff values obtained from our quantum mechanical model.
The Newton equation for an electron moving in the $z$-axis can be written: 
\begin{eqnarray}
\label{newton}
\nonumber
\ddot{z}(t)&=&-\nabla_z V_l(z,t)\\
&=&-(E_{dc}+E_{L}(t))
\end{eqnarray}
Eq.~(\ref{newton}) is solved under the following conditions: (i) the electron starts
its movement with zero velocity at the origin at time $t = t_i$ ($t_i$ is known
as ionization time), i.e., $z(t_i) = 0$ and $\dot{z}(t_i) = 0$; (ii) when the electric field reverses its direction, the
electron returns to its initial position (i.e., recollides with the
ionic core) at a later time, $t = t_r$ ($t_r$ is also known as recollision
time), i.e., $z(t_r)=0$. By modifying the values of the ionization time $t_i$, it is possible to
compute the classical electron trajectories and to numerically calculate
the times $t_r$ where the recollision phenomenon takes place (for numerical details see e.g.~\cite{ciappicpc2014}). Furthermore,
once the ionization time $t_i$ is fixed, the electron trajectory
is completely determined. The electron kinetic energy, as a function of the ionization $t_i$ or recombination $t_r$ times, can be then calculated by using $E_k(t_j)=\dot{z}(t_j)^{2}/2$ where $j = i$ for ionization or $j = r$ for recollision. Finally, the harmonic order on the ionization ($t_i$) and recollision times ($t_r$), is obtained from $n = (E_k(t_j) + W)/\omega$ with $j=i,r$ for ionization or recombination times, respectively.

\section{Results}
\label{sec:3}

We start by computing spectra using Eq.~(\ref{hhg}) employing a laser pulse of $n_{cy}=10$ cycles long at a wavelength $\lambda=800$ nm (30 fs of total time duration) and for different values of the laser peak amplitude $E_0$ and the static electric field $E_{dc}$. In Fig.~1 we use a peak electric field $E_0=10$ GV m$^{-1}$, meanwhile the spectra of Fig.~2 are for a $E_0=20$ GV m$^{-1}$. The different panels correspond different values of the static field $E_{dc}$, namely panels (a) $E_{dc}=-0.4$ GV m$^{-1}$, panels (b) $E_{dc}=0$ and panels (c) $E_{dc}=+2$ GV m$^{-1}$, respectively.  
\vspace{0.75cm}
\begin{figure}[h]
\resizebox{0.45\textwidth}{!}{\includegraphics{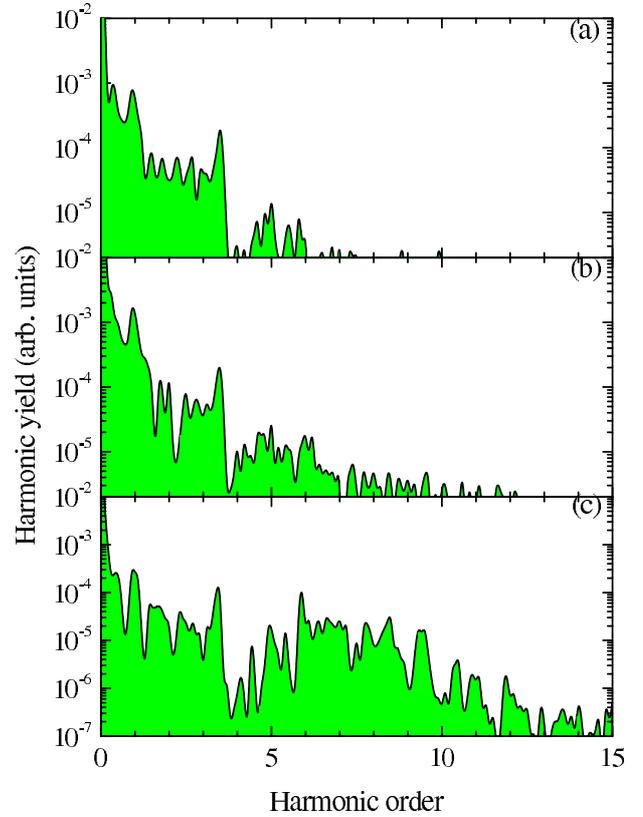}
}
%\vspace{5cm}       % Give the correct figure height in cm
%\vspace{-.75cm}
\caption{(color online) HHG spectra as a function of harmonic order for a metal (Au) nanotip using a sin-squared shaped laser pulse with 10 cycles of total duration, $\protect\lambda=800$ nm and $E_0=10$ GV m$^{-1}$. Panel a) $E_{dc}=-0.4$ GV m$^{-1}$; panel b) $E_{dc}=0$ and panel c) $E_{dc}=+2$ GV m$^{-1}$.}
\label{fig:1}       % Give a unique label
\end{figure}

As in the case of atoms a clear harmonic cutoff can be seen at $n_c\approx5$ (equivalent to a photon energy of 7.75 eV and a wavelength $\lambda=160$ nm) for $E_0=10$ GV m$^{-1}$ (Fig.~1(b)) and one at $n_c\approx 10$ (equivalent to a photon energy of 15.5 eV and a wavelength $\lambda=80$ nm) for $E_0=20$ GV m$^{-1}$ (Fig.~2(b)). Negative values of the static electric field have a minor influence (see Figs.~1(a) and 2(a)), but, on the other hand, when positive values are used, a clear extension in the harmonic cutoff can be observed (see Figs.~1(c) and 2(c)). Harmonics of the order $n_c\approx 15$ (equivalent to a photon energy of 23.25 eV and a wavelength $\lambda=53$ nm) can be achieved (Fig.~2(c)). In addition, for $E_{dc}>0$ an increase in the harmonic yield is observed (a similar behaviour was predicted by shorter pulses, see~\cite{ciappiNanotip}) and this feature could be experimentally exploited considering a larger harmonic signal would be much easier to detect.

Next we employ the classical model described in Section 2.2 in order to study the electron kinetic energy and to characterize the harmonic cutoff. 
\vspace{0.5cm}
\begin{figure}[h]
\resizebox{0.45\textwidth}{!}{\includegraphics{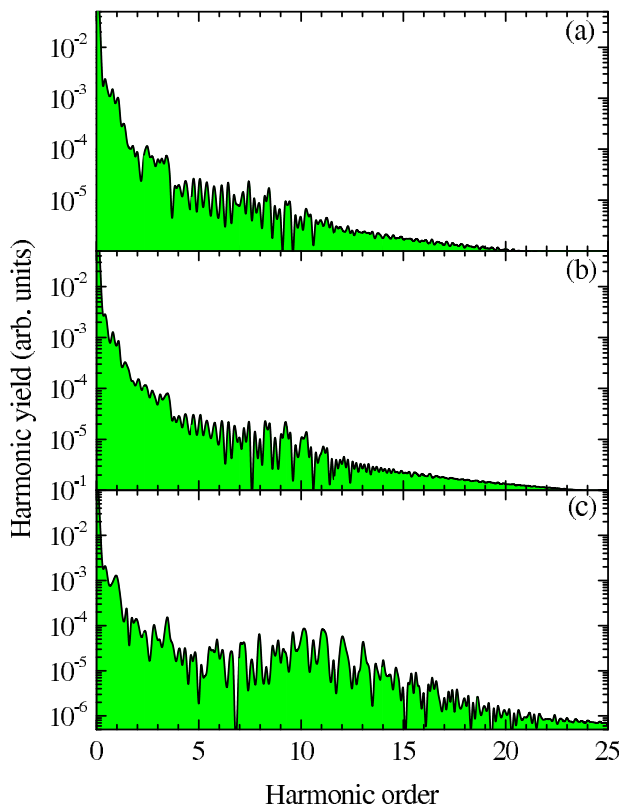}
}
%\vspace{5cm}       % Give the correct figure height in cm
%\vspace{-.75cm}
\caption{Idem Fig.~1 for a $E_0=20$ GV m$^{-1}$.}
\label{fig:2}       % Give a unique label
\end{figure}

Figs.~3 and 4 we plot the dependence of the harmonic order on the ionization ($t_i$) (green (light gray) circles) and recollision times ($t_r$) (red (dark gray) circles) following the prescriptions of the classical approach and using the same laser parameters as in the quantum simulations. Fig.~3 is the classical counterpart of Fig.~1, meanwhile Fig.~4 is the classical counterpart of Fig.~2, respectively. 
We observe that the classical calculations confirm the modifications of the harmonic cutoff due to the incorporation of the static field $E_{dc}$, in reasonable agreement with our full quantum mechanical model. The deviation in harmonic cutoff from the conventional simple man's model is more pronounced for the case of the positive $E_{dc}$, as predicted by the quantum mechanical model. Furthermore, it seems that the positive $E_{dc}$ not just extend the harmonic cutoff but it modifies some of the electron trajectories. This can be clearly noticed when we compare the Fig.~4(b) with Fig.~4(c). As was observed for short pulses~\cite{ciappiNanotip}, for negative $E_{dc}$ the classical model predicts a larger cutoff in comparison to the quantum mechanical calculations. This can be explained by investigating the transition amplitude of the individual electron trajectories. It seems that, herein the quantum transition amplitude of those trajectories contributing to the larger cutoff values is very small. As a result they do not show up in the harmonic spectra. However, we need a more exhaustive investigation to fully explore how the presence of a $E_{dc}$ affects the HHG spectra. 

\vspace{0.75cm}
\begin{figure}[h]
\resizebox{0.475\textwidth}{!}{\includegraphics{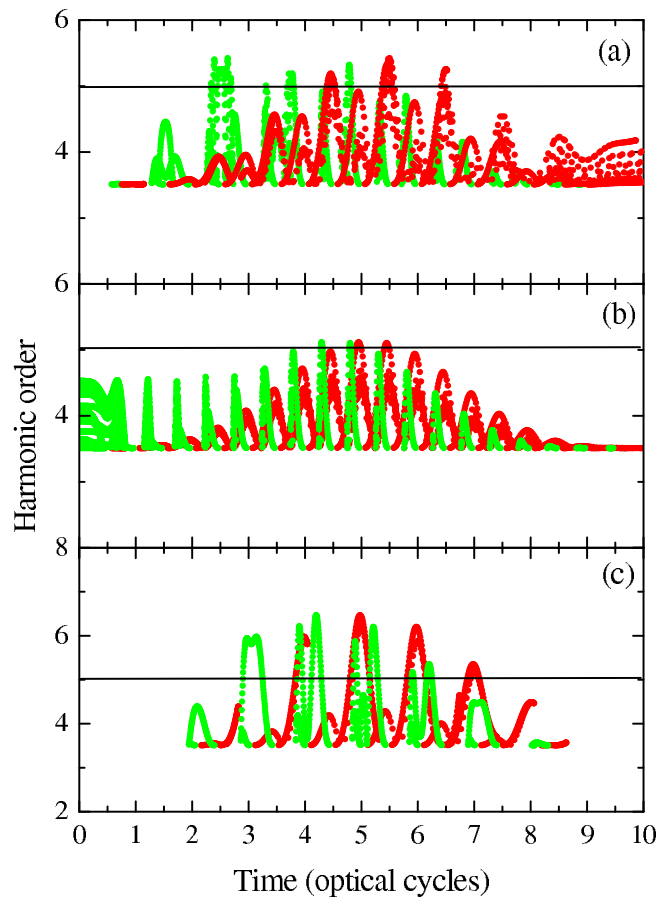}
}
%\vspace{5cm}       % Give the correct figure height in cm
%\vspace{-.75cm}
\caption{(Color online) Total energy of the electron (in terms
of the harmonic order $n$) obtained from Newton's second law and plotted as a
function of the ionization time $t_i$ (green (light gray) circles) or the recollision
time $t_r$ (red (dark gray) circles) for a peak electric field $E_0=10$ GV m$^{-1}$. The solid line parallel to the $x$-axis at around $n_c\approx 5$ shows the harmonic cutoff predicted by the conventional simple man's model (see text). Panel (a): $E_{dc}=-0.4$ GV m$^{-1}$, panel (b): $E_{dc}=0$ and panel (c): $E_{dc}=+2$ GV m$^{-1}$.}
\label{fig:3}       % Give a unique label
\end{figure}

Furthermore, by comparing the conventional classical calculations (Figs.~3(b) and 4(b)) with those when a $E_{dc}$ is present (independently of its sign) we observe a clear break in the symmetry of the plots around the middle of the laser pulse ($n_{cy}=5$ for this case). Particularly, this is seen for the electron kinetic energy as a function of the recombination time (red (dark gray) circles). A follow up contribution will address this feature.


%\begin{figure*}
%% Use the relevant command for your figure-insertion program
%% to insert the figure file. See example above.
%% If not, use
%\vspace*{5cm}       % Give the correct figure height in cm
%\end{figure*}

\section{Conclusions and Outlook}
\label{sec:4}

In conclusion, we have extended previous predictions of high-order harmonics generation directly from metal sharp tips. Our quantum mechanical model is used in order to compute the HHG yield, now employing a longer laser pulse at the Ti:Sapphire wavelength ($\lambda=800$ nm). We observed a noticeable extension of the HHG cutoff when positive static fields are employed. 

%\vspace{0.75cm}
\begin{figure}[h]
\resizebox{0.475\textwidth}{!}{\includegraphics{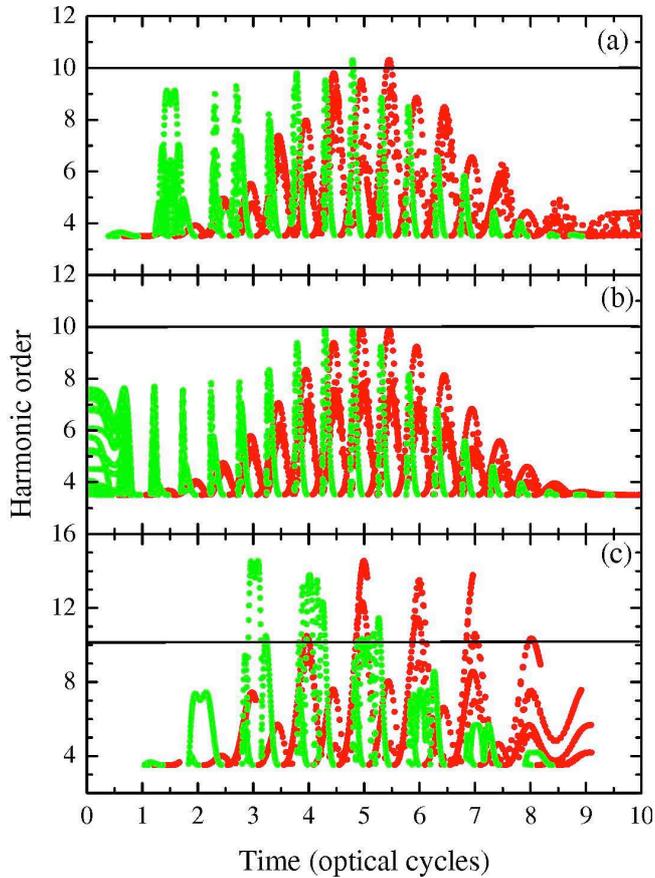}
}
%\vspace{5cm}       % Give the correct figure height in cm
%\vspace{-.75cm}
\caption{(Color online) Total energy of the electron (in terms
of the harmonic order $n$) obtained from Newton's second law and plotted as a
function of the ionization time $t_i$ (green (light gray) circles) or the recollision
time $t_r$ (red (dark gray) circles) for a peak electric field $E_0=20$ GV m$^{-1}$. The solid line parallel to the $x$-axis at around $n_c\approx 10$ shows the harmonic cutoff predicted by the conventional simple man's model (see text). Panel (a): $E_{dc}=-0.4$ GV m$^{-1}$, panel (b): $E_{dc}=0$ and panel (c): $E_{dc}=+2$ GV m$^{-1}$.}
\label{fig:4}       % Give a unique label
\end{figure}

As before we neglected any spatial variation of the (plasmonically) enhanced field, but this aspect might become important for other experimental conditions, which will be considered in a future work. The classical simulations, including the static electric field, confirm the quantum mechanical results in a higher degree of accuracy. The simulations presented in this work corroborate that the harmonic emission from metal nanotips can be a reliable alternative for the generation of coherent radiation at the UV to XUV range.

\begin{acknowledgement}
We acknowledge the financial support of the MICINN
projects (FIS2008-00784 TOQATA, FIS2008-06368-C02-01,
and FIS2010-12834), ERC Advanced Grant QUAGATUA and OSYRIS, the
Alexander von Humboldt Foundation, the Hamburg Theory
Prize (M.L.), and the DFG Cluster of Excellence Munich Center for Advanced Photonics. This research has been partially supported
by Fundaci\`o Privada Cellex. J.A.P.-H. acknowledges support
from the Spanish MINECO through the Consolider Program
SAUUL (CSD2007-00013) and research project FIS2009-
09522, from Junta de Castilla y Le\'on through the Program
for Groups of Excellence (GR27), and from the ERC Seventh
Framework Programme (LASERLAB-EUROPE, Grant No.
228334). This work was made possible in part by a grant of high performance computing
resources and technical support from the Alabama Supercomputer Authority. We thank Peter Hommelhoff and Michael Kr\"uger for useful comments and suggestions.
\end{acknowledgement}

%%as required. Don't forget to give each section
%%and subsection a unique label (see Sect.~\ref{sec:1}).
%%
%% For one-column wide figures use
%\begin{figure}
%% Use the relevant command for your figure-insertion program
%% to insert the figure file.
%% For example, with the option graphics use
%\resizebox{0.75\textwidth}{!}{%
%  \includegraphics{leer.eps}
%}
%% If not, use
%%\vspace{5cm}       % Give the correct figure height in cm
%\end{figure}
%%
%% For two-column wide figures use
%\begin{figure*}
%% Use the relevant command for your figure-insertion program
%% to insert the figure file. See example above.
%% If not, use
%\vspace*{5cm}       % Give the correct figure height in cm
%\end{figure*}
%

%\begin{table}
%\begin{tabular}{lll}
%\hline\noalign{\smallskip}
%\end{tabular}
%% Or use
%\vspace*{5cm}  % with the correct table height
%\end{table}
%
% BibTeX users please use
%\bibliographystyle{plain}
%\bibliography{plasmonics}

\begin{thebibliography}{}
\bibitem{McPherson1987} A. McPherson, G. Gibson, H. Jara, U. Johann, T. S. Luk, I. A. McIntyre, K. Boyer, C. K. Rhodes, J. Opt. Soc. Am. B {\bf 4}, 595 (1987)
\bibitem{Huillier1991} A. L'Huillier, K. J. Schafer, K. C. Kulander, J. Phys. B {\bf 24}, 3315 (1991)
\bibitem{antoine} P. Antoine, A. L'Huillier, M. Lewenstein, Phys. Rev. Lett. {\bf 77} 1234 (1996)
%\bibitem{corkumnat} P. B. Corkum, F. Krausz, Nat. Phys. {\bf 3}, 381 (2007)
\bibitem{Scrinzi2004} A. Scrinzi, T. Westerwalbesloh, U. Kleineberg, U. Heinzmann, M. Drescher, F. Krausz, Nature {\bf 427}, 817 (2004)
\bibitem{hhg1khz} S. H. Chew, F. S{\"u}{\ss}mann, C. Sp{\"a}th, A. Wirth, J. Schmidt, S. Zherebtsov, A. Guggenmos, A. Oelsner, N. Weber, J. Kapaldo, et al., App. Phys. Lett. {\bf 100}, 051904 (2012)
\bibitem{hhg3khz} M. Schultze, E. Goulielmakis, M. Uiberacker, M. Hofstetter, J. Kim, D. Kim, F. Krausz, U. Kleineberg, New J. Phys. {\bf 9}, 243 (2007)
\bibitem{amelle} M. Krebs, S. H{\"a}drich, S. Demmler, J. Rothhardt, A. Za{\"i}r, L. Chipperfield, J. Limpert, A. T{\"u}nnermann, Nat. Phot. {\bf 7}, 555 (2013)
\bibitem{manfred_rev} M. Lein, J. Phys. B {\bf 43}, R135 (2007)
\bibitem{pacer} S. Baker, J. S. Robinson, C. A. Haworth, H. Teng, R. A. Smith, C. C. Chiril\u{a}, M. Lein, J. G. Tisch, J. P. Marangos, Science {\bf 312}, 424 (2006)
\bibitem{rabitt} S. Haessler, J. Caillat, W. Boutu, C. Giovanetti-Teixeira, T. Ruchon, T. Auguste, Z. Diveki, P. Breger, A. Maquet, B. C., R. Ta{\"i}ıeb, et al., Nat. Phys. {\bf 6}, 200 (2010)
\bibitem{olga1} O. Smirnova, Y. Mairesse, S. Patchkovskii, N. Dudovich, D. Villeneuve, P. Corkum, M. Y. Ivanov, Proc. Natl. Acad. Sci. USA {\bf 106}, 16556 (2009)
\bibitem{olga2} O. Smirnova, Y. Mairesse, S. Patchkovskii, N. Dudovich, D. Villeneuve, P. Corkum, M. Y. Ivanov, Nature (London) {\bf 460}, 972 (2009)
\bibitem{mairesse} Y. Mairesse, A. de Bohan, L. J. Frasinski, H. Merdji, L. C. Dinu, P. Monchicourt, P. Breger, M. Kova\u{c}ev, R. Ta{\"i}eb, B. Carr{\'e}, et al., Science {\bf 302}, 1540 (2003)
\bibitem{power} E. P. Power, A. M. March, F. Catoire, E. Sistrun, K. Krushelnick, P. Agostini, L. F. DiMauro, Nat. Phot. {\bf 4}, 352 (2010)
\bibitem{peter2011} M. Kr\"uger, M. Schenk, P. Hommelhoff, Nature {\bf 475}, 78 (2011)
\bibitem{klingnature} S. Zherebtsov, T. Fennel, J. Plenge, E. Antonsson, I. Znakovskaya, A. Wirth, O. Herrwerth, F. S\"u{\ss}mann, C. Peltz, I. Ahmad, et al., Nat. Physics {\bf 7}, 656 (2011)
\bibitem{ganeev1} C. Hutchison, R. A. Ganeev, T. Witting, F. Frank, W. A. Okell, J. W. G. Tisch, J. P. Marangos, Opt. Lett. {\bf 37}, 2064 (2012)
\bibitem{ganeev2} R. A. Ganeev, T. Witting, C. Hutchison, F. Frank, P. V. Redkin, W. A. Okell, D. Y. Lei, T. Roschuk, S. A. Maier, J. P. Marangos, et al., Phys. Rev. A {\bf 85}, 015807 (2012)
\bibitem{peter2012} M. Kr\"uger, M. Schenk, M. F\"orster, P. Hommelhoff, J. Phys. B {\bf 45}, 074006 (2012)
\bibitem{peter2010} M. Schenk, M. Kr\"uger, P. Hommelhoff, Phys. Rev. Lett. {\bf 105}, 257601 (2010)
\bibitem{herink} G. Herink, D. R. Solli, M. Gulde, C. Ropers, Nature {\bf 483}, 190 (2012)
\bibitem{yalunin} S. V. Yalunin, M. Gulde, C. Ropers, Phys. Rev. B {\bf 84}, 195426 (2011)
\bibitem{michaelNJP} M. Kr\"uger, M. Schenk, P. Hommelhoff, G. Watcher, C. Lemell, J. Burgd{\"o}rfer, New J. Phys. {\bf 14}, 085019 (2012)
\bibitem{watcherPRB} G. Watcher, C. Lemell, J. Burgd{\"o}rfer, M. Schenk, M. Kr{\"u}ger, P. Hommelhoff, Phys. Rev. B {\bf 86}, 085019 (2012)
\bibitem{ropersAdP} S. V. Yalunin, G. Herink, D. R. Solli, M. Kr\"uger, P. Hommelhoff, M. Diehn, A. Munk, C. Ropers, Ann. Phys. (Berlin) {\bf 525}, L12 (2013)
\bibitem{ghimireexp} S. Ghimire, A. D. DiChiara, E. Sistrunk, P. Agostini, L. F. DiMauro, D. A. Reis, Nat. Phys. {\bf 7}, 138 (2011)
\bibitem{ghimiretheory} S. Ghimire, A. D. DiChiara, E. Sistrunk, G. Ndabashimiye, U. B. Szafruga, A. Mohammad, P. Agostini, L. F. DiMauro, D. A. Reis, Phys. Rev. A {\bf 85}, 043836 (2012)
\bibitem{kim} S. Kim, J. Jin, Y.-J. Kim, I.-Y. Park, Y. Kim, S.-W. Kim, Nature {\bf 453}, 757 (2008)
\bibitem{muhl} P. M\"uhlschlegel, H.-J. Eisler, O. J. F. Martin, B. Hecht, D. W. Pohl, Science {\bf 308}, 1607 (2005)
\bibitem{schuck} P. J. Schuck, D. P. Fromm, A. Sundaramurthy, G. S. Kino, W. E. Moerner, Phys. Rev. Lett. {\bf 94}, 017402 (2005)
\bibitem{husakou} A. Husakou, S.-J. Im, J. Herrmann, Phys. Rev. A {\bf 83}, 043839 (2011)
\bibitem{yavuz} I. Yavuz, E. A. Bleda, Z. Altun, T. Topcu, Phys. Rev. A {\bf 85}, 013416 (2012)
\bibitem{ciappi2012} M. F. Ciappina, J. Biegert, R. Quidant, M. Lewenstein, Phys. Rev. A {\bf 85}, 033828 (2012)
\bibitem{tahir2012} T. Shaaran, M. F. Ciappina, M. Lewenstein, Phys. Rev. A {\bf 86}, 023408 (2012)
\bibitem{ciappi_opt} M. F. Ciappina, S. S. A{\'c}imovi{\'c}, T. Shaaran, J. Biegert, R. Quidant, M. Lewenstein, Opt. Exp. {\bf 20}, 26261 (2012)
\bibitem{ciappiAdP} T. Shaaran, M. F. Ciappina, M. Lewenstein, Ann. Phys. (Berlin) {\bf 525}, 97 (2013)
\bibitem{miloAdP} B. Feti\'c, K. Kalajd\u{z}i\'c, D. B. Milo\u{s}evi\'c, Ann. Phys. (Berlin) {\bf 525}, 107 (2013)
\bibitem{joseprl2013} J. A. P\'erez-Hern\'andez, M. F. Ciappina, M. Lewenstein, L. Roso, A. Za\"ir, Phys. Rev. Lett. {\bf 110}, 053001 (2013)
\bibitem{yavuz2013} I. Yavuz, Phys. Rev. A {\bf 87}, 053815 (2013)
\bibitem{luo2013} J. Luo, Y. Li, Z. Wang, Q. Zhang, P. Lu, J. Phys. B {\bf 46}, 145602 (2013)
\bibitem{PoP} L. Feng, M. Yuan, T. Chu, Phys. Plasmas {\bf 20}, 122307 (2013)
\bibitem{ropersnat} M. Sivis, M. Duwe, B. Abel, C. Ropers, Nature {\bf 485}, E1 (2012)
\bibitem{Kimreply} S. Kim, J. Jin, Y.-J. Kim, I.-Y. Park, Y. Kim, S.-W. Kim, Nature {\bf 485}, E2 (2012)
\bibitem{sivis2013} M. Sivis, M. Duwe, B. Abel, C. Ropers, Nat. Phys. {\bf 9}, 304 (2013)
\bibitem{klingspie} F. S\"u{\ss}mann, M. F. Kling, Proc. of SPIE {\bf 8096}, 80961C (2011)
\bibitem{klingprb} F. S\"u{\ss}mann, M. F. Kling, Phys. Rev. B {\bf 84}, 121406(R) (2011)
\bibitem{lastkling} Y.-Y. Yang, A. Scrinzi, A. Husakou, Q.-G. Li, S. L. Stebbings, F. S\"u{\ss}mann, H.-J. Yu, S. Kim, E. R\"uhl, J. Herrmann, et al., Opt. Exp. {\bf 21}, 2195 (2013)
\bibitem{peter2006} P. Hommelhoff, Y. Sortais, A. Aghajani-Talesh, M. A. Kasevich, Phys. Rev. Lett. {\bf 96}, 077401 (2006)
\bibitem{peter2006a} P. Hommelhoff, C. Kealhofer, M. A. Kasevich, Phys. Rev. Lett. {\bf 97}, 247402 (2006)
\bibitem{Ropers2007} C. Ropers, D. R. Solli, C. P. Schulz, C. Lienau, T. Elsaesser, Phys. Rev. Lett. {\bf 98}, 043907 (2007)
\bibitem{Barwick2007} B. Barwick, C. Corder, J. Strohaber, N. Chandler-Smith, C. Uiterwaal, H. Batelaan, New J. Phys. {\bf 9}, 142 (2007)
\bibitem{Yanagisawa2009} H. Yanagisawa, C. Hafner, P. Don\'a, M. Kl\"ockner, D. Leuenberger, T. Greber, M. Hengsberger, J. Osterwalder, Phys. Rev. Lett. {\bf 103}, 257603 (2009)
\bibitem{Bormann2010} R. Bormann, M. Gulde, A. Weismann, S. V. Yalunin, C. Ropers, Phys. Rev. Lett. {\bf 105}, 147601 (2010)
\bibitem{Park2012} D. J. Park, B. Piglosiewicz, S. Schmidt, H. Kollmann, M. Mascheck, C. Lienau, Phys. Rev. Lett. {\bf 109}, 244803 (2012)
\bibitem{ciappiNanotip} M. F. Ciappina, J. A. P\'erez-Hern\'andez,, T. Shaaran, M. Lewenstein, M. Kr\"uger, P. Hommelhoff, Phys. Rev. A {\bf 89} 013409 (2014)
\bibitem{keitel} M. Protopapas, C. H. Keitel, P. L. Knight, Rep. Prog. Phys. {\bf 60}, 389 (1997)
\bibitem{schafer} K. J. Schafer, K. C. Kulander, Phys. Rev. Lett. {\bf 78}, 638 (1997)
\bibitem{sfa} M. Lewenstein, P. Balcou, M. Y. Ivanov, A. L'Huillier, P. B. Corkum, Phys. Rev. A {\bf 49}, 2117 (1994)
\bibitem{corkum} P. B. Corkum, Phys. Rev. Lett {\bf 71}, 1994 (1993)
\bibitem{ciappicpc2014} M. F. Ciappina, J. A. P\'erez-Hern\'andez, M. Lewenstein, Comp. Phys. Commun. {\bf 185}, 398 (2014)

%Author, \textit{Book title} (Publisher, place year) page numbers
% etc
\end{thebibliography}

\end{document}